\documentclass[conference]{IEEEtran}
\usepackage{amsmath,amsfonts}
\usepackage{amsthm}
\usepackage{algorithmic}
\usepackage{algorithm}
\usepackage{array}
\usepackage[caption=false,font=footnotesize,labelfont=sf,textfont=sf]{subfig}
\usepackage{textcomp}
\usepackage{stfloats}
\usepackage{url}
\usepackage{verbatim}
\usepackage{cite}
\usepackage{color}
\usepackage{cases}
\usepackage[pdftex]{graphicx}
\usepackage{epstopdf}
\newcommand{\RNum}[1]{\uppercase\expandafter{\romannumeral #1\relax}}
\newcounter{foo}
\newtheorem{theorem}{Theorem}

\DeclareMathAlphabet\mathbfcal{OMS}{cmsy}{b}{n}

\def\BibTeX{{\rm B\kern-.05em{\sc i\kern-.025em b}\kern-.08em
    T\kern-.1667em\lower.7ex\hbox{E}\kern-.125emX}}
\begin{document}

\title{On Analysis of Superimposed Pilot in Multi-User Massive MIMO with Massive Connectivity\\
\thanks{Identify applicable funding agency here. If none, delete this.}
}

\author{
\IEEEauthorblockN{
Shuxiao Ye\IEEEauthorrefmark{1},    
Xianchao Zhang\IEEEauthorrefmark{2}, and
Neng Ye\IEEEauthorrefmark{3}}
\IEEEauthorblockA{\IEEEauthorrefmark{1}School of Information and Electronics, Beijing Institute of Technology, Beijing, China.}
\IEEEauthorblockA{\IEEEauthorrefmark{2}Provincial Key Laboratory of Multimodal Perceiving and Intelligent Systems, Jiaxing University, Jiaxing, China.}
\IEEEauthorblockA{\IEEEauthorrefmark{3}School of Cyberspace Science and Technology, Beijing Institute of Technology, Beijing, China. }
\IEEEauthorblockA{Email: shuxiaoye@bit.edu.cn, zhangxianchao@zjxu.edu.cn, ianye@bit.edu.cn}}

\maketitle

\begin{abstract}

    The simultaneous transmission of numerous users presents substantial challenges due to the inherent trade-off between channel estimation and information transmission in multi-user multiple-input multiple-output (MIMO) system. 
    In this paper, we explore the use of the superimposed pilot (SP) scheme to tackle the large transmitting users, where the number of users may exceed the coherent time. 
    SP scheme incorporates both transmitted data and noise in the channel estimation process, which is significant different from the counterpart of RP scheme. 
    We provide an in-depth analysis of the interaction between interference caused by channel estimation errors and noise.
    We then derive the explicit expression for the scaling law of the mutual information lower bound (MILB) in relation to the number of users and the levels of transmitted power. 
    Besides, the optimal power allocation between pilots and data transmission is also derived analytically.
    The analytical results demonstrate that the SP scheme significantly improves performance compared to traditional RP scheme in our consider case. 
    Numerical results are also presented to validate our theoretical derivations.    
    
\end{abstract}


\section{Introduction}
With the rapid advancement of the ubiquitous and widespread deployment in next generation communications, wireless communication networks are confronting the significant challenge of managing concurrent access from a massive number of devices \cite{10492466}. 
In this circumstance, massive multiple-input multiple-output (MIMO) technology, especially in a multi-user setup, with its inherent advantage of spatial freedom, has emerged as a crucial solution to resolve the tension between capacity and the number of connections \cite{10379539}. 
The full harassment of the spatial diversity gain offered by multi-user MIMO, it is essential to rely on accurate acquisition of channel state information (CSI).

In single-user MIMO communication, channel estimation is performed using a set of known pilots, which are transmitted prior to the actual data transmission \cite{svantesson_capacity_2005}. 
It is challenging to analytically determine the mutual information between the channel input and output, making it difficult to compute the precise channel capacity in this context.
Besides, while investing excessive resources in training can improve channel estimation accuracy, it concurrently consumes valuable data transmission resources, resulting in a significant reduction in overall system capacity. 
To address these issues, Hassibi introduces the concept of using the mutual information lower bound (MILB) as an alternative metric, providing a novel approach to training-based capacity analysis in MIMO systems \cite{hassibi_how_2003}.
Furthermore, Coldrey and Bohlin refine the study by focusing on the optimization trade-off between the pilot and data transmission, with respect to MILB of MIMO fading channel \cite{coldrey_training-based_2007}.

Different from single-user MIMO, multi-user MIMO systems must ensure the communication quality for all user links.
To accommodate it, multi-user MIMO systems typically employ orthogonal pilot frequency allocation strategies for channel estimation.
However, these methods require training resources proportional to the number of users, which becomes impractical as the demand for massive user access continues to grow \cite{ngo_energy_2013}. 
To mitigate this issue, the innovative approach of employing non-orthogonal pilot sequences has been proposed to reduce training overhead \cite{liu_massive_2018}. 
Based on this approach, \cite{yuan_fundamental_2018} systematically establishes a quantitative relationship between the channel estimation error under non-orthogonal frequency pilots and the optimal number of serviceable users.

On the other hand, the regular pilot (RP) overhead in the training phase can be avoided through pilot-data superposition, referred to as superimposed pilot (SP) scheme \cite{zhang_superimposed_2016}. 
This approach allows users to adopt pilot sequences of the same length as the data, significantly increasing the number of users that can maintain orthogonality \cite{verenzuela_spectral_2018}. 
In this scheme, the average spectral efficiency of the users is investigated by considering combining methods such as maximum ratio (MR) \cite{haritha_superimposed_2024}, zero-forcing (ZF) \cite{zhou_optimized_2024}, and minimum mean square error (MMSE) \cite{duran_superimposed_2024}. 
These methods aim to improve the user-specific signal-to-interference-plus-noise ratio (SINR) by optimizing the power allocation between the user's data and pilot \cite{zhou_power_2022}.
However, when the number of concurrently transmitting users exceeds the number of available coherent time, it becomes impossible to preserve the orthogonality of the pilots across users, even with the superimposed pilot scheme. 

In this paper, we investigate a multi-user MIMO system where the number of transmitted users exceeds the coherent time. 
Specifically, we focus on the application of the SP scheme with respect to the MILB.
We firstly characterize the interactive between the channel estimation error, transmitted pilots and messages, and noise.
Then, the explicit expression for the scale law of MILB in the non-asymptotic case is derived.
Besides, the optimal power allocation trade off between the transmitted pilot and transmitted messages for the SP is demonstrated with the analytical derivation.
Additionally, we also evaluated the performance of the RP and compared it with the SP.
Finally, numerical results are presented to validate our theoretical findings.

\section{System Model}

\subsection{Signal Model}

We consider an uplink transmission of massive MIMO systems, where $K$ single-antenna users transmit signals to a base station (BS) through the same channel. 
The signal transmitted by the \( k \)-th terminal is represented by \( \mathbf{x}_k \in \mathbb{C}^{1 \times L} \) with power $P$, which satisfies the power constraint as \( \mathbb{E}[\mathbf{x}_k \mathbf{x}_k^{H}] \leq L \).
The BS equips $N$ received antennas, and the channel between $k$th terminal to BS is denoted by $\mathbf{h}_k = (h_{1,k},\dots,h_{N,k})^{T} \in \mathcal{CN}(\mathbf{0},\mathbf{I}_{N})$. 
We assume the block-fading, i.e., the channel keeps invariant within the coherent time L.
The corresponding channel for a frame of received signal with $L$ time at the BS is given by
\begin{equation}
    \mathbf{Y} 
    = \sum_{k=1}^K \mathbf{h}_{k} \sqrt{P} \mathbf{x}_k^T + \mathbf{N}
    = \sqrt{P} \mathbf{H} \mathbf{X}^T + \mathbf{N},
\end{equation}
where $\mathbf{H} = [\mathbf{h}_{1}, \mathbf{h}_{2}, \cdots, \mathbf{h}_{K}] \in \mathbb{C}^{N \times K} $ is the channel matrix for the combination of $K$ users, the matrix \( \mathbf{X} = [\mathbf{x}_1^T, \mathbf{x}_2^T, \cdots, \mathbf{x}_K^T]^T \in \mathbb{C}^{K \times L} \) contains the transmitted signals of all \( K \) users, and \( \mathbf{N} \in \mathbb{C}^{N \times L} \) is the Gaussian noise matrix with each element \( n_{i,j} \) follows the distribution \( n_{i,j} \sim \mathcal{CN}(0, \sigma^2) \).

\begin{figure}[t]
    \flushleft 
    \hspace{0.25in}
    \includegraphics[width=2.5in]{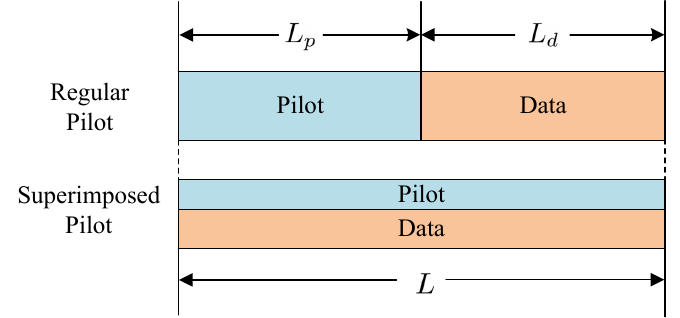}
    \caption{The structure of transmission signal under RP and SP schemes.}
    \label{fig:Signal_structure}
\end{figure}   

\subsection{Superimposed Pilot}

For the SP scheme, the users transmit the pilot and data simultaneously.
The pilot and the data are superimposed over the coherent time, i.e., the transmitted signal of the \( k \)-th user is given by
\begin{equation}
    \mathbf{x}_k = \sqrt{\alpha} \boldsymbol{\phi}_k + \sqrt{(1-\alpha)} \mathbf{s}_k,
\end{equation}
where \( \alpha \in (0,1) \) is the power allocation factor, and \( \boldsymbol{\phi}_k \in \mathbb{C}^{1 \times L} \) is the pilot of the \( k \)-th user, and \( \mathbf{s}_k \in \mathbb{C}^{1 \times L} \) is the data of the \( k \)-th user, which satisfies \( \mathbf{s}_k \sim \mathcal{CN}(\mathbf{0}, \mathbf{I}_L) \).

Due to the large number of users exceeding the coherent time ,i.e., $K > L$, the pilot assigned to each user must be non-orthogonal to each other.
In this context, we consider a maximum-Welch-bound-equality (MWBE) sequence \cite{4784839}, where the pilot of $k$th user is given by
\begin{equation}
    \boldsymbol{\phi}_k = [1, e^{- 2\pi j \frac{k-1}{K}}, \cdots, e^{- 2\pi j \frac{k-1}{K}(L-1)} ] \in \mathbb{C}^{1 \times L}.
\end{equation}
The $\boldsymbol{\Phi} = [\boldsymbol{\phi}_1, \boldsymbol{\phi}_2, \cdots, \boldsymbol{\phi}_K ] \in \mathbb{C}^{K \times L}$ is defined as the pilot matrix, which contains the pilot symbols of all users, and satisfies
\begin{equation}
    \boldsymbol{\Phi} \boldsymbol{\Phi}^H = \mathbf{R}_{\phi}, \space \boldsymbol{\Phi}^H \boldsymbol{\Phi} = K \mathbf{I}_L,
\end{equation}
where the element in $(\mathbf{R}_{\phi})_{i,j}$ is given by $\sum_{n=1}^{L} e^{-2 \pi j \frac{i-j}{K} (n-1)}$. 

The received signal at BS is given by
\begin{equation}
    \begin{aligned}
        \mathbf{Y} = \sqrt{\alpha P} \mathbf{H} \boldsymbol{\Phi} + \sqrt{(1 - \alpha) P} \mathbf{H} \mathbf{S} + \mathbf{N},
    \end{aligned}
\end{equation}
where \( \mathbf{Y} \in \mathbb{C}^{N \times L} \), and \( \mathbf{S} = [\mathbf{s}_1, \mathbf{s}_2, \cdots, \mathbf{s}_K ] \in \mathbb{C}^{K \times L} \) is the data signal matrix of all users.

Here, we employ the linear minimum mean square error (MMSE) channel estimate and consider the data as noise. 
In this case, the channel estimate process matrix is given by
\begin{equation}
        \mathbf{W} 
        = \dfrac{\sqrt{\alpha P} }{P K + \sigma^2} \boldsymbol{\Phi}^{H}, \\
\end{equation} 
where $\mathbf{R}_{H} = \mathbb{E}\left[ \mathbf{H}^{H} \mathbf{H} \right] = K \mathbf{I}_{L}$.
And the estimated channel matrix is given by
\begin{equation} \label{eq:channel_estimate}
    \mathbf{\hat{H}}
    = \mathbf{Y} \mathbf{W}
    = \dfrac{\sqrt{\alpha P} }{P K + \sigma^2} \mathbf{Y} \boldsymbol{\Phi}^{H}.
\end{equation}
After channel estimation, the estimated channel is considered as the true channel.
We define $\mathbf{\tilde{H}}$ as the channel estimate error, which satisfies $\mathbf{H} = \mathbf{\hat{H}} + \mathbf{\tilde{H}}$.
Then, receiver would extract the pilots based on the estimated channel.
Consequently, the received data $\mathbf{Y}_d$ could be express as 
\begin{equation}
    \begin{aligned}
        \mathbf{Y}_d 
        =& \mathbf{Y} - \sqrt{\alpha P} \hat{\mathbf{H}}\boldsymbol{\Phi} \\
        =& \sqrt{(1 - \alpha) P} \hat{\mathbf{H}} \mathbf{S} + \underbrace{\tilde{\mathbf{H}} \left( \sqrt{\alpha P} {\mathbf{\Phi}} + \sqrt{(1 - \alpha) P} \mathbf{S} \right)  + \mathbf{N}}_{\mathbf{V}}.  \\
    \end{aligned}
\end{equation}
Here we use the $\mathbf{V}$ to denote the combination of interference caused by channel estimate error and noise, which satisfies $\mathbb{E} \left[ \mathbf{S} \mathbf{V}^{H} \right] = 0$.
It is known that ``worse-case" noise for the additive channel follows the Gaussian distribution and independent of transmission data \cite{hassibi_how_2003}.

Considering the known and observed at both the input and output of the channel, we could obtain the mutual information as  
\begin{equation}
    \begin{aligned}
        & I\left( \mathbf{Y}, \boldsymbol{\Phi} ; \mathbf{S} \right) \\
        \geq & I\left( \mathbf{Y}_d, \hat{\mathbf{H}} ; \mathbf{S} \right) \\
        \geq & I\left( \mathbf{Y}_{d} ; \mathbf{S} | \hat{\mathbf{H}} \right) + I\left( \hat{\mathbf{H}}, \boldsymbol{\Phi}; \mathbf{S} | \boldsymbol{\Phi} \right), \\    
    \end{aligned}
\end{equation}
where $I\left( \hat{\mathbf{H}}, \boldsymbol{\Phi}; \mathbf{S} | \boldsymbol{\Phi} \right) = I\left( \hat{\mathbf{H}}; \mathbf{S} \right) \geq 0$.
In this case, the channel capacity is lower-bound by the conditional mutual information as
\begin{equation}
    I\left( \mathbf{S}; \mathbf{Y}_d | \hat{\mathbf{H}}\right) 
    = \mathbb{E}_{\mathbf{H}} \left[ \log{ \operatorname{det} \left( \dfrac{(1 - \alpha) P}{\sigma^2_V} \mathbf{\hat{H}} \mathbf{\hat{H}}^H + \mathbf{I}_{N} \right) } \right] .
\end{equation}

\subsection{Regular Pilot}

For the RP, the transmission is divided into two phases, i.e., pilot transmission phase and data transmission phase.
In the pilot transmission phase, users only transmit pilot which occupies $L_p$ coherent time.
Non-orthogonal pilots are generated in the same manner as in SP scheme. 
The pilot of $k$-th is given by
\begin{equation}
    \boldsymbol{\phi}_k^{\rm{RP}} = [1, e^{- 2\pi j \frac{k-1}{K}}, \cdots, e^{- 2\pi j \frac{k-1}{K}(L_p-1)} ] \in \mathbb{C}^{1 \times L_p}.
\end{equation}
Since the pilot in RP is shorter than the that in SP, it will increase the correlation between pilots and reduce the accuracy of channel estimate.
The matrix $\boldsymbol{\Phi}^{\rm{RP}} = [{\boldsymbol{\phi}_1^{\rm{RP}}}^T, {\boldsymbol{\phi}_2^{\rm{RP}}}^T, \cdots, {\boldsymbol{\phi}_K^{\rm{RP}}}^T ]^T \in \mathbb{C}^{K \times L_p} $ is defined as the pilot matrix which contains the pilots of all users, and satisfies
\begin{equation}
    \boldsymbol{\Phi}^{\rm{RP}} \left(\boldsymbol{\Phi}^{\rm{RP}}\right)^H = \mathbf{R}_{\phi^{\rm{RP}}}, \space \left(\boldsymbol{\Phi}^{\rm{RP}}\right)^H \boldsymbol{\Phi}^{\rm{RP}} = K \mathbf{I}_{L_p},
\end{equation}
where the element in $\mathbf{R}_{\phi^{\rm{RP}}}$ is given by $\sum_{n=1}^{L} e^{-2 \pi j \frac{i-j}{K} (n-1)}$. 

In the data transmission phase, users transmit the data in the remains $L_d = L - L_p$ time.
The entire transmission signal of the \( k \)-th user can be represented as
\begin{equation}
    \mathbf{x}_k = [{\boldsymbol{\phi}_k^{\rm{RP}}}, {\mathbf{s}_k^{\rm{RP}}}],
\end{equation}
where $\mathbf{s}_k^{\rm{RP}} \in \mathbb{C}^{1 \times L_{d} }$ is the encoded signal of the \( k \)-th user, which satisfies \( \mathbf{s}_k^{\rm{RP}} \sim \mathcal{CN}(\mathbf{0}, \mathbf{I}_{L_d }) \).

The received signal can be expressed as
\begin{equation}
    \mathbf{Y}^{\rm{RP}} = \sqrt{P} \mathbf{H} {[{\boldsymbol{\Phi}^{\rm{RP}}}, {\mathbf{S}^{\rm{RP}}}]} + \mathbf{N},
\end{equation}
where \( \mathbf{S}^{\rm{RP}} = [{\mathbf{s}_1^{\rm{RP}}}^T, {\mathbf{s}_2^{\rm{RP}}}^T, \cdots, {\mathbf{s}_K^{\rm{RP}}}^T ]^T \in \mathbb{C}^{K \times (L-L_p)} \) is the encoded signal matrix of all users.

Let $\mathbf{Y}_{p}^{\rm{RP}}$ be the first $L_p$ columns of $\mathbf{Y}$, which is represented received signal obtained in pilot transmission phase, and $\mathbf{Y}_{d}^{\rm{RP}}$ be remains $L_d$ columns of $\mathbf{Y}$, which is the received signal obtained in data transmission phase.
In this case, the channel estimate process matrix is given by
\begin{equation}
    \mathbf{W}^{\rm{RP}} 
    = \dfrac{\sqrt{P}}{PK + \sigma^2} {\boldsymbol{\Phi}^{\rm{RP}}}^{H}.
\end{equation}
The MMSE estimated channel matrix $\hat{\mathbf{H}}^{\rm{RP}}$ is given by
\begin{equation}
    \mathbf{\hat{H}}^{\rm{RP}} 
    = \mathbf{Y}_{p}^{\rm{RP}} \mathbf{W}^{\rm{RP}} 
    = \dfrac{\sqrt{P}}{PK + \sigma^2} \mathbf{Y}_{p}^{\rm{RP}} {\boldsymbol{\Phi}^{\rm{RP}}}^{H}.
\end{equation}
We similarly use $\mathbf{\tilde{H}}^{\rm{RP}}$ as the channel estimate error, which satisfies $\mathbf{\tilde{H}}^{\rm{RP}} = \mathbf{H} - \hat{\mathbf{H}}^{\rm{RP}}$.

Based on the estimated channel matrix, we can decompose the received signal and have
\begin{equation}
    \mathbf{Y}_{d}^{\rm{RP}} 
    = \sqrt{P} \mathbf{\hat{H}}^{\rm{RP}} \mathbf{S}^{\rm{RP}} + \underbrace{\sqrt{P} \mathbf{\tilde{H}}^{\rm{RP}} \mathbf{S}^{\rm{RP}} + \mathbf{N}_d}_{\mathbf{V}^{\rm{RP}}},
\end{equation}
where $\mathbf{V}^{\rm{RP}}$ satisfies $\mathbb{E} \left[ \mathbf{S} \left( \mathbf{V}^{\rm{RP}} \right) ^{H} \right] = 0$. 
Similarly, the MILB for RP is given by
\begin{equation}
    \begin{aligned}
        &I\left( \mathbf{S}^{\rm{RP}}; \mathbf{Y}_d^{\rm{RP}} | \hat{\mathbf{H}}^{\rm{RP}} \right) \\
        =& \dfrac{L_d}{L} \mathbb{E}_{\mathbf{H}}\left(\log{ \operatorname{det} \left( \dfrac{P}{\sigma^2_{\mathbf{V}^{\rm{RP}}}} \mathbf{\hat{H}}^{\rm{RP}} \left( \mathbf{\hat{H}}^{\rm{RP}} \right)^{H} + \mathbf{I}_{N} \right) } \right). \\
    \end{aligned}    
\end{equation}

\section{Performance Analysis}

\subsection{Performance of Superimposed Pilot}

Since the estimated channel matrix $\mathbf{\hat{H}}$ is obtained by linearly processing the received signal.
According to the distributions of noise, transmitted data and channel fading, we note that the received signal $\mathbf{Y}$ is a random matrix with zero mean with covariance matrix
\begin{equation}
    \begin{aligned}
        &\mathbb{E} \left[\mathbf{Y}^{H} \mathbf{Y}\right] \\
        =& \mathbb{E}\left[ \alpha P \boldsymbol{\Phi}^{H} \mathbf{H}^{H} \mathbf{H} \boldsymbol{\Phi} + (1 - \alpha) P \mathbf{S}^{H} \mathbf{H}^{H} \mathbf{H} \mathbf{S} + N \sigma^2 \mathbf{I}_{L} \right] \\
        =& N \left( P K + \sigma^2 \right) \mathbf{I}_{L}       
    \end{aligned}
\end{equation} 
Therefore, $\mathbf{Y}$ can be equivalently expressed as
\begin{equation} \label{eq:Y_equivalent}
    \begin{aligned}
        \mathbf{Y}
        =& \mathbf{G} \boldsymbol{\Sigma}_{G} \\
        =& \left( P K + \sigma^2 \right)^{\frac{1}{2}} \mathbf{G},
    \end{aligned}
\end{equation}
where $\mathbf{G} \in \mathbb{C}^{N \times L}$ is a random matrix with the elements independently drawn from $\mathcal{CN}(0,1)$, and $\boldsymbol{\Sigma}_{G} = \mathbb{E} \left[\mathbf{Y}^{H} \mathbf{Y}\right]^{\frac{1}{2}}$.
Furthermore, combining \eqref{eq:Y_equivalent} and \eqref{eq:channel_estimate}, we could obtain the following equivalent
\begin{equation} \label{eq:Co_H_hat_equivalent}
    \begin{aligned}
        \mathbf{\hat{H}} \mathbf{\hat{H}}^{H} 
        =& \mathbf{G} \boldsymbol{\Sigma}_{G} \mathbf{W} \mathbf{W}^{H} {\boldsymbol{\Sigma}_{G}}^{H} \mathbf{G}^{H}\\
        =& \dfrac{\alpha P K}{P K + \sigma^2} \mathbf{G} \mathbf{G}^{H}.
    \end{aligned} 
\end{equation}
Meanwhile, since the MMSE channel estimation error $\mathbf{\tilde{H}}$ is independent of the estimation value $\mathbf{\hat{H}}$, we can get that the variance of the channel estimation error $\sigma_{\mathbf{\tilde{H}}}^2$ satisfies
\begin{equation}
    \begin{aligned}
        \sigma_{\mathbf{\tilde{H}}}^2 =& 
        \dfrac{1}{N K} \mathbb{E} \left\| \mathbf{\tilde{H}} \right\|_F^2 \\
        =& \dfrac{1}{N K} \operatorname{tr} \left( \mathbf{R}_{H} -  \mathbb{E} \left[ \mathbf{\hat{H}} \mathbf{\hat{H}}^{H} \right]\right) \\
        =& 1 - \dfrac{\alpha P L}{P K + \sigma^2},
    \end{aligned}
\end{equation}
where $\operatorname{tr}(\cdot)$ denotes the trace of the matrix.

Now, we focus on the interference caused by channel estimation error and noise.
Since channel estimate is determined by both the current noise and channel realization, it would arise significant correlation among the terms within $\mathbf{V}$.
\begin{theorem} \label{thm:IaN_variance}
    The variance of element in $\mathbf{V}$ is given by
    \begin{equation}
        \begin{aligned}
            \sigma_{\mathbf{V}}^2 
            =& \sigma^2 + (1 - \alpha) P K \left( 1 - \dfrac{\alpha P}{P K + \sigma^2} L  \right) 
            \\ & + \alpha P K \left( 1 - \dfrac{\alpha P}{P K + \sigma^2} K \right) - 2 \dfrac{\alpha P}{P K + \sigma^2} K \sigma^2
            \\ & + 2 \dfrac{\alpha^2 (1 - \alpha) P^3}{(P K + \sigma^2)^2} K^2 L - 2 \dfrac{\alpha (1-\alpha) P^2}{P K + \sigma^2} K^2 .
        \end{aligned}
    \end{equation}
\end{theorem}
\begin{IEEEproof}
    Please see the appendix.
\end{IEEEproof}

By concluding \eqref{eq:Co_H_hat_equivalent} and Theorem \eqref{thm:IaN_variance}, we can derive the MILB for SP as
\begin{equation} \label{eq:I_lower_bound_exact}
    \begin{aligned}
        I\left( \mathbf{S}; \mathbf{Y}_{d} | \hat{\mathbf{H}}\right)  
        =& \mathbb{E}_{\mathbf{H}}\left(\log{ \operatorname{det} \left( \dfrac{ (1 - \alpha) P \sigma_{\mathbf{G}}^2}{\sigma^2_V} \mathbf{\bar{G}} {\mathbf{\bar{G}}}^{H} + \mathbf{I}_{N} \right) } \right) \\
        =& L \int_{0}^{\infty} \log \left( 1 + \rho \lambda  \right) f_{L}(\lambda) \operatorname{d} \lambda, \\
    \end{aligned}
\end{equation}
where $\sigma^{2}_{\mathbf{G}} = \frac{\alpha P K}{ P K + \sigma^2}$, $f_L(\lambda)$ is the probability density function of the eigenvalue of $\mathbf{\bar{G}}^{H} \mathbf{\bar{G}}$, which draws from Wishart distribution $\mathcal{W}_{L}^{(N)}(N, \mathbf{I}_{L})$, is given by
\begin{equation}
    f_{L}(\lambda) = \dfrac{1}{L} \sum_{m=0}^{L-1} \dfrac{m!}{\left(m + N - L\right)!} \left(\rm{La}_{m}^{(N-L)}(\lambda) \right)^2 \lambda^{N-L} e^{-\lambda}, 
\end{equation}
and $\rm{La}_{m}^{(N-L)}(\lambda)$ is the $m$-th order generalized Laguerre polynomial with
\begin{equation}
    \rm{La}_{m}^{r}(\lambda) = \sum_{k=0}^{m} (-1)^{i} \binom{m + r}{m - k} \dfrac{\lambda^{k}}{k !}.
\end{equation}
Besides, $\rho = \dfrac{\alpha (1 - \alpha)}{ \gamma} P^2 K$ is given by
\begin{equation}
    \begin{aligned}
        \gamma =
        & - \frac{2 K^2 L P^3}{\sigma^2 + K P} \alpha^3\\
        & + \left( \frac{2 K^2 L P^3 }{\sigma^2 + K P} + K L P^2 + K^2 P^2 \right) \alpha^2 \\
        & + \left(- K L P^2 - 2 K^2 P^2 - 2 K P \sigma^2 \right)  \alpha \\
        & + \left( \sigma^2 + K P \right)^2. \\
    \end{aligned}
\end{equation}

We aim to find the optimal value of $\alpha$ that maximizes the MILB for SP.
In fact, only $\rho$ in \eqref{eq:I_lower_bound_exact} depends on $\alpha$.
Besides, the MILB for SP is monotonically increased with respect to $\rho$.
Thus, we can obtain the optimal value of $\alpha$ by differentiating $\rho$.
\begin{theorem} \label{thm:alpha_optimal}
    The optimal value of $\alpha = \alpha^{*}$ that maximum the MILB for SP, is given by the real root of the equation
    \begin{equation}
        \begin{aligned}
            g(\alpha) =& 
            - \frac{2 K^2 L P^3}{\sigma^2 + K P} \alpha^4 + 2 \frac{2 K^2 L P^3}{\sigma^2 + K P} \alpha^3 \\
            &- \left( \frac{2 K^2 L P^3 }{\sigma^2 + K P} - K^2 P^2 - 2 K P \sigma^2 \right) \alpha^2 \\
            &- 2 \left( \sigma^2 + K P \right)^2 \alpha + \left( \sigma^2 + K P \right)^2 , 
        \end{aligned}
    \end{equation}
    which falls within the interval $(0, 1)$.
\end{theorem}
\begin{IEEEproof}
    Considering the derivative of $\rho$ is given by
    \begin{equation}
        \dfrac{\partial \rho}{\partial \alpha} = P^2 K \dfrac{g(\alpha)}{\gamma^2},
    \end{equation}
    and the fact that $\alpha$ in constraint in $(0, 1)$. 
\end{IEEEproof}

\subsection{Performance of Regular Pilot}

For the RP, the channel estimate operates only during the pilot transmission phase.
Similarly, we can express the received signal in pilot transmit transmission phase as
\begin{equation}
    \mathbf{Y}_{p} = \left( P K + \sigma^2  \right)^{\frac{1}{2}} \mathbf{G}^{\rm{RP}} ,
\end{equation}
where $\mathbf{G}^{\rm{RP}} \in \mathbb{C}^{N \times L_p}$ with each element drawing from $\mathcal{CN}(0, 1)$, and $\boldsymbol{\Sigma}_{\mathbf{G}^{\rm{RP}}} = \left(P \boldsymbol{\Phi}^{H} \boldsymbol{\Phi} + \sigma \mathbf{I}_{L_p}\right)^{\frac{1}{2}} $.
Therefore, we can also obtain the equivalent for estimated channel under RP as
\begin{equation} \label{eq:Co_H_hat_equivalent_regular}
    \begin{aligned}
        \mathbf{\hat{H}}^{\rm{RP}} \left( \mathbf{\hat{H}}^{\rm{RP}} \right)^{H}
        =& \dfrac{P K}{P K + \sigma^2} \mathbf{G}^{\rm{RP}} (\mathbf{G}^{\rm{RP}})^{H} .
    \end{aligned}
\end{equation} 
The channel estimate error $\mathbf{\tilde{H}}^{\rm{RP}}$ is also independent of the estimated channel $\mathbf{\hat{H}}^{\rm{RP}}$, i.e., $\mathbb{E}\left[ \mathbf{H} \left(\mathbf{\tilde{H}}^{\rm{RP}}\right)^{H} \right] = \mathbf{0}$.
The covariance of channel estimate error $\sigma^2_{\mathbf{\tilde{H}}^{\rm{RP}}}$ under RP is given by
\begin{equation}
    \begin{aligned}
        \sigma_{\mathbf{\tilde{H}}^{\rm{RP}}}^{2} 
        &= \dfrac{1}{N K} \mathbb{E} \left\| \mathbf{\tilde{H}}^{\rm{RP}} \right\|_F^2 \\
        &= \dfrac{1}{N K} \operatorname{tr} \left( \mathbf{R}_{H} - \mathbb{E} \left[ \mathbf{\hat{H}}^{\rm{RP}} \left( {\mathbf{\hat{H}}^{\rm{RP}}} \right)^{H} \right] \right) \\
        &= 1 - \dfrac{P L_p}{P K + \sigma^2} .
    \end{aligned}
\end{equation}

In the data transmission phase, the noise matrix $\mathbf{N}_{d}$ is independent of the estimated channel $\mathbf{\hat{H}}^{\rm{RP}}$ and transmission data $\mathbf{S}^{\rm{RP}}$.
Thus, we could easily obtain variance of $\mathbf{V}^{\rm{RP}}$ as
\begin{equation} \label{eq:V_RP_variance}
    \begin{aligned}
        \sigma_{\mathbf{V}^{\rm{RP}}}^2 
        &= \dfrac{1}{N L_d} \mathbb{E} \left\| \mathbf{V}^{\rm{RP}} \right\|_F^2 \\
        &= \dfrac{1}{N L_d} \operatorname{tr} \left( P \mathbb{E} \left[\mathbf{\tilde{H}}^{\rm{RP}} \mathbf{R}_{\mathbf{S}^{\rm{RP}}} \left(\mathbf{\tilde{H}}^{\rm{RP}}\right)^{H}\right] + \mathbf{R}_{\mathbf{N}_d} \right) \\
        &= P K \sigma_{\mathbf{\tilde{H}}^{\rm{RP}}}^{2} + \sigma^2 ,
    \end{aligned}
\end{equation}
where $\mathbf{R}_{\mathbf{N}_d} = L_d \mathbf{I}_{N}$ is the noise covariance matrix in data transmission phase, and $\mathbf{R}_{\mathbf{S}^{\rm{RP}}} = L_d \mathbf{I}_{K}$ is the correlation matrix of transmitted data.

Now, we could evaluate the MILB by combining \eqref{eq:V_RP_variance} and \eqref{eq:Co_H_hat_equivalent_regular} and obtain
\begin{equation} \label{eq:I_lower_bound_regular}
    \begin{aligned}
        &I\left( \mathbf{S}^{\rm{RP}}; \mathbf{Y}_{d}^{\rm{RP}} | \hat{\mathbf{H}}^{\rm{RP}}\right)  \\ 
        =& \dfrac{L_d}{L} \mathbb{E}_{\mathbf{H}}\left[ \log{ \operatorname{det} \left( \dfrac{P \sigma^2_{\mathbf{G}^{\rm{RP}}}}{\sigma^2_{\mathbf{V}^{\rm{RP}}}} \mathbf{G}^{\rm{RP}} \left(\mathbf{G}^{\rm{RP}}\right)^{H} + \mathbf{I}_{N} \right) } \right] \\
        =& \dfrac{L - L_p}{L} L_p \int_{0}^{\infty}  \log \left( 1 + \rho^{\rm{RP}} \lambda f_{L_p}(\lambda)  \right) \operatorname{d} \lambda ,
    \end{aligned}
\end{equation}
where $\sigma^2_{\mathbf{G}^{\rm{RP}}} = \dfrac{P K}{P K + \sigma^2}$, $f_{L_p}$ is defined as $f_{L}$ in previous, and $\rho^{\rm{RP}}$ satisfies
\begin{equation}
    \rho^{\rm{RP}}
    = \dfrac{P^2 K}{P^2 K (K-L_p) + \sigma^4 + 2 P K \sigma^2} .
\end{equation}
However, $L_d$ is involved into the coefficients of the integral, $f_{L_p}(\lambda)$, and $\rho^{\rm{RP}}$. 
As a result, it is impossible to derivative the explicit expression of the optimal value with respect to $L_d$ in \eqref{eq:I_lower_bound_regular}.
In this circumstance, we obtain the optimal $L_d$ by an exhaustive search over $\{1, \cdots, L-1\}$.

\section{Numerical Results and Discussion}

In this section, we present the numerical results to illustrate the theories described in this paper.
We compare the scale law of MILBs interactive with the number of transmitted users and the different levels of transmitted power.
Unless otherwise specified, we set the simulation parameters as follows:
$L=30$, $N=60$, and $\sigma^2 = 1$.

\begin{figure}[t]
    \centering
    \includegraphics[width=\linewidth]{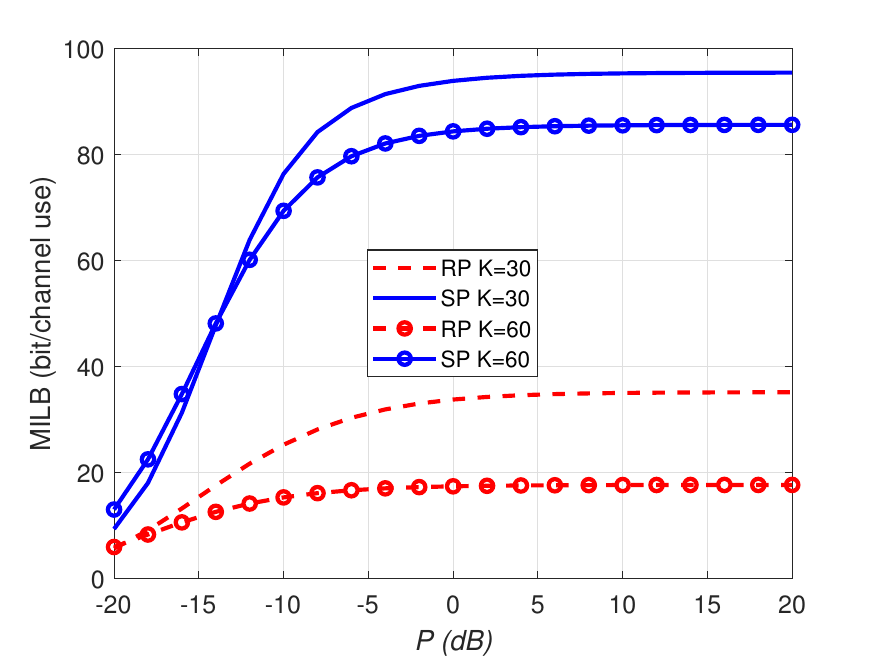}
    \caption{The MILB versus the levels of transmitted power $P$ under different users $K$.}
    \label{fig:MILB_vs_Power}
\end{figure}    

\begin{figure}[t]
    \centering
    \includegraphics[width=\linewidth]{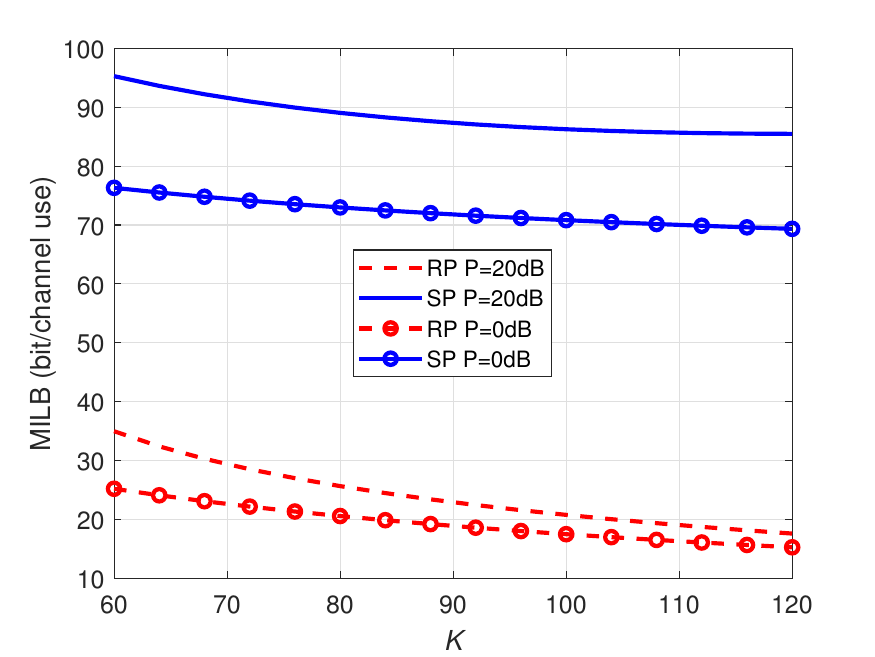}
    \caption{The MILB versus the number of transmitted users $K$ under different power levels $P$.}
    \label{fig:MILB_vs_K}
\end{figure}    

Fig. \ref{fig:MILB_vs_Power} illustrates the MILBs versus the levels of transmitted power with the number of transmitted users $K = 10$ and $K = 40$.
The SP has a great performance than RP in terms of the MILB.
However, due to the use of non-orthogonal pilots in channel estimation, the variance of the channel estimation error cannot be reduced to zero. 
As a result, the MILB reaches a performance floor at high transmitted power levels
In this case, the SP facilitates the longer pilots than RP, which may lead to the higher channel estimate accuracy and enable the channel capacity.

Fig. \ref{fig:MILB_vs_K} shows MILBs against the number of transmitted users with the levels of transmit power $P = 20dB$ and $P = 0dB$.
The SP has also demonstrated the significant performance improvement compared to RP.
Besides, it can observe that MILB is proportional inversely to number of transmitted users.
That means that there is still trade off between the number of transmitted users and the MILB under SP, which is already verified for RP in \cite{yuan_fundamental_2018}.

\section{Conclusion}

In this paper, we investigated the SP with respect to MILB under the multi-user MIMO system where the number of transmitted users is larger than the coherent time.
We firstly derived the explicit expression for the scale law of MILB in the non-asymptotic case.
Besides, we also presented the detail interactive among the channel estimation error, transmitted pilots and messages, and noise.
The optimal power allocation trade off between the transmitted pilot and transmitted messages for the SP was also demonstrated.
Similarly, we also evaluated the performance of the RP and compared it with the SP.
Since the large transmitted users had to employ the non-orthogonal pilots, the SP facilitated the lower interference sequence and significantly improved the performances.

\appendix

We could decompose the covariance matrix of the interference caused by channel estimate error and noise as
\begin{equation}
    \begin{aligned}
        \sigma_{\mathbf{V}}^2 =
        &\dfrac{1}{N L} \operatorname{tr} \mathbb{E} [ \mathbf{V} \mathbf{V}^{H} ] \\
        =&\dfrac{1}{N L}  \operatorname{tr} \big( \mathbf{R}_1 + \mathbf{R}_2 + \mathbf{R}_3 \\
        & + \mathbf{R}_4 + \mathbf{R}_4^{H} + \mathbf{R}_5 + \mathbf{R}_5^{H} + \mathbf{R}_6 + \mathbf{R}_6^{H} \big) ,\\
    \end{aligned}
\end{equation}
where the trace of $\mathbf{R}_1, \mathbf{R}_2, \mathbf{R}_3$ could be directly obtained as
\begin{equation}
    \begin{aligned}
        &\begin{aligned}
            \operatorname{tr}(\mathbf{R}_1) = \operatorname{tr} \left(\mathbb{E}[\mathbf{N} \mathbf{N}^{H}]\right) =& L N \sigma^2 \\
        \end{aligned}\\
        &\begin{aligned}
            \operatorname{tr}(\mathbf{R}_2) 
            =& \operatorname{tr} \left( (1 - \alpha) P \mathbb{E}[\tilde{\mathbf{H}} \mathbf{S} {\mathbf{S}}^{H} \tilde{\mathbf{H}}^{H}]  \right)\\
            =&  (1 - \alpha) P L K N \left( 1 - \dfrac{\alpha P}{ P K + \sigma^2} L  \right) \\
        \end{aligned}\\
        &\begin{aligned}
            \operatorname{tr}(\mathbf{R}_3)
           =& \operatorname{tr} \left( \alpha P \mathbb{E}[\tilde{\mathbf{H}} \boldsymbol{\Phi} {\boldsymbol{\Phi}}^{H} \tilde{\mathbf{H}}^{H}] \right) \\
           =& \alpha P \operatorname{tr} \left( {\boldsymbol{\Phi}}^{H} \mathbb{E}\left[ \tilde{\mathbf{H}}^{H}  \tilde{\mathbf{H}}  \right] \boldsymbol{\Phi} \right) \\
           =& \alpha P K L N \left( 1 - \dfrac{\alpha P}{P K + \sigma^2} K \right).
        \end{aligned}
    \end{aligned}
\end{equation}

For the remaining terms, we need to detail analyze the interactive between $\mathbf{N}$, $\mathbf{S}$, and $\mathbf{\tilde{H}}$. For $\mathbf{R}_4$, we have
\begin{equation}
    \begin{aligned}
        \operatorname{tr}(\mathbf{R}_4) =& \operatorname{tr} \left( \sqrt{\alpha P} \mathbb{E} \left[ \tilde{\mathbf{H}} \boldsymbol{\Phi} \mathbf{N}^{H}\right]\right) \\
        =& \sqrt{\alpha P} \operatorname{tr} \left( \mathbb{E} \left[ \left( \mathbf{H} - \hat{\mathbf{H}}\right) \boldsymbol{\Phi} \mathbf{N}^{H}\right]\right) \\
        =& - \sqrt{\alpha P} \operatorname{tr} \left( \mathbb{E} \left[ \hat{\mathbf{H}} \boldsymbol{\Phi} \mathbf{N}^{H}\right]\right) \\
        =& - \sqrt{\alpha P} \operatorname{tr} \left( \mathbb{E} \left[ \dfrac{\sqrt{\alpha P}}{P K + \sigma^2} \mathbf{Y} \boldsymbol{\Phi}^{H} \boldsymbol{\Phi} \mathbf{N}^{H}\right]\right) \\
        =& - \dfrac{\alpha P K}{P K + \sigma^2} \operatorname{tr} \left( \mathbb{E} \left[ \mathbf{Y} \mathbf{N}^{H}\right]\right) \\
        =& - \dfrac{\alpha P K}{P K + \sigma^2} \operatorname{tr} \left( \mathbb{E} \left[ \mathbf{N} \mathbf{N}^{H}\right]\right) \\
        =& - \dfrac{\alpha P }{P K + \sigma^2} K L N \sigma^2 .\\
    \end{aligned}
\end{equation}
And for $\mathbf{R}_5$, we have
\begin{equation}
    \begin{aligned}
        \operatorname{tr}(\mathbf{R}_5) =& \operatorname{tr} \left( \sqrt{(1-\alpha) P} \mathbb{E} \left[ \mathbf{N} \mathbf{S} \tilde{\mathbf{H}}^{H} \right] \right) \\
        =& \operatorname{tr} \left( \sqrt{(1-\alpha) P} \mathbb{E} \left[ \mathbf{N} \mathbf{S} \left( \mathbf{H} - \hat{\mathbf{H}}\right)^{H} \right] \right) \\
        =& -\sqrt{(1-\alpha) P} \operatorname{tr} \left( \mathbb{E} \left[ \mathbf{N} \mathbf{S} \hat{\mathbf{H}}^{H} \right] \right) \\
        =& -\sqrt{(1-\alpha) P} \dfrac{\sqrt{\alpha P}}{PK + \sigma^2} \operatorname{tr} \left( \mathbb{E} \left[ \mathbf{N} \mathbf{S} \boldsymbol{\Phi} \mathbf{Y}^{H} \right] \right) \\
        =& -\dfrac{\sqrt{\alpha (1-\alpha)}P}{PK + \sigma^2} \operatorname{tr} \left( \mathbb{E} \left[ \mathbf{N} \mathbf{S} \boldsymbol{\Phi} \mathbf{N}^{H} \right. \right. \\ 
        & \left. \left. +  \sqrt{(1-\alpha)P} \mathbf{N} \mathbf{S} \boldsymbol{\Phi} \mathbf{S}^{H} \right] \right)  \\
        =& 0 . \\
    \end{aligned}
\end{equation}
The term $\mathbf{R}_6$ satisfies as
\begin{equation}
    \begin{aligned}
        & \operatorname{tr}(\mathbf{R}_6) \\
        =& \operatorname{tr} \left( \sqrt{\alpha (1-\alpha)} P \mathbb{E} \left[ \tilde{\mathbf{H}} \mathbf{S} \boldsymbol{\Phi}^{H} \tilde{\mathbf{H}}^{H} \right] \right) \\
        =& \sqrt{\alpha (1-\alpha)} P \operatorname{tr} \left( \mathbb{E}\left[ \boldsymbol{\Phi}^{H} \tilde{\mathbf{H}}^{H} \tilde{\mathbf{H}} \mathbf{S} \right] \right) \\
        =& \sqrt{\alpha (1-\alpha)} P \operatorname{tr} \left( \mathbb{E}\left[ \boldsymbol{\Phi}^{H} \left( \mathbf{H} - \hat{\mathbf{H}}\right)^{H} \left( \mathbf{H} - \hat{\mathbf{H}}\right) \mathbf{S} \right] \right) \\
        =& \sqrt{\alpha (1-\alpha)} P \operatorname{tr} \left( \mathbb{E}\left[ \boldsymbol{\Phi}^{H} \mathbf{H}^{H} \mathbf{H} \mathbf{S} - \boldsymbol{\Phi}^{H} \hat{\mathbf{H}}^{H} \mathbf{H} \mathbf{S} \right. \right.
        \\ & \left. \left.  - \boldsymbol{\Phi}^{H} \mathbf{H}^{H} \hat{\mathbf{H}} \mathbf{S} + \boldsymbol{\Phi}^{H} \hat{\mathbf{H}}^{H} \hat{\mathbf{H}} \mathbf{S} \right] \right) \\
        =& \sqrt{\alpha (1-\alpha)} P \operatorname{tr} \left( \mathbb{E}\left[ \dfrac{\alpha P}{(P K + \sigma^2)^2} \boldsymbol{\Phi}^{H} \boldsymbol{\Phi} \mathbf{Y}^{H} \mathbf{Y} \boldsymbol{\Phi}^{H} \mathbf{S} \right. \right.
        \\ & \left. \left.  - \dfrac{\sqrt{\alpha P}}{P K + \sigma^2} \boldsymbol{\Phi}^{H} \mathbf{H}^{H} \mathbf{Y} \boldsymbol{\Phi}^{H} \mathbf{S} - \dfrac{\sqrt{\alpha P}}{P K + \sigma^2} \boldsymbol{\Phi}^{H} \boldsymbol{\Phi} \mathbf{Y}^{H} \mathbf{H} \mathbf{S}  \right] \right) .\\
    \end{aligned}
\end{equation}
The first term follows
\begin{equation}
    \begin{aligned}
        &\operatorname{tr} \left( \mathbb{E}\left[ \dfrac{\alpha P}{(P K + \sigma^2)^2} \boldsymbol{\Phi}^{H} \boldsymbol{\Phi} \mathbf{Y}^{H} \mathbf{Y} \boldsymbol{\Phi}^{H} \mathbf{S} \right] \right) \\
        =& \dfrac{\alpha P}{(P K + \sigma^2)^2} K \operatorname{tr} \left( \mathbb{E}\left[ \mathbf{Y}^{H} \mathbf{Y} \boldsymbol{\Phi}^{H} \mathbf{S} \right] \right) \\
        =& \dfrac{\alpha P}{(P K + \sigma^2)^2} K \operatorname{tr} \left( \mathbb{E}\left[ \left( \sqrt{\alpha (1 - \alpha)} P \boldsymbol{\Phi}^{H} \mathbf{H}^{H} \mathbf{H} \mathbf{S} \right. \right. \right.
        \\ & \left. \left. \left. + \sqrt{\alpha (1 - \alpha)} P \mathbf{S}^{H} \mathbf{H}^{H} \mathbf{H} \boldsymbol{\Phi}  \right) \boldsymbol{\Phi}^{H} \mathbf{S} \right] \right) \\
        =& \dfrac{\alpha P}{(P K + \sigma^2)^2} K \sqrt{\alpha (1 - \alpha)} P N 
        \\ & \cdot \left( \operatorname{tr} \left( \mathbb{E}\left[ \boldsymbol{\Phi}^{H} \mathbf{S} \boldsymbol{\Phi}^{H} \mathbf{S} \right] \right)  + \operatorname{tr}\left( \mathbb{E} \left[ \boldsymbol{\Phi}^{H} \mathbf{S} \mathbf{S}^{H} \boldsymbol{\Phi} \right] \right) \right) \\
        =& \dfrac{\alpha \sqrt{\alpha (1 - \alpha)} P^2}{(P K + \sigma^2)^2} K^2 L^2 N .\\
    \end{aligned}
\end{equation}
The last equation is noted for $\mathbb{E}\left[ \boldsymbol{\Phi}^{H} \mathbf{S} \boldsymbol{\Phi}^{H} \mathbf{S} \right] = \mathbf{0}_{L}$. 
This is due to the fact that each element in $\mathbb{E}\left[ \boldsymbol{\Phi}^{H} \mathbf{S} \boldsymbol{\Phi}^{H} \mathbf{S} \right]$ is cumulative from the squares of the complex numbers, and $(a + bj)^2 = a^2 - b^2 + 2abj$, where $a, b$ is independently followed by $\mathcal{CN}(0,\sigma^2/2)$.
Thus, we have $\mathbb{E}\left[ (a + bj)^2 \right] = 0$ and achieve the desired result.
The second term follows
\begin{equation}
    \begin{aligned}
        &\operatorname{tr} \left( \mathbb{E}\left[ \dfrac{\sqrt{\alpha P}}{P K + \sigma^2} \boldsymbol{\Phi}^{H} \mathbf{H}^{H} \mathbf{Y} \boldsymbol{\Phi}^{H} \mathbf{S} \right] \right) \\
        =& \dfrac{\alpha P}{P K + \sigma^2} \operatorname{tr} \left( \mathbb{E}\left[\boldsymbol{\Phi}^{H} \mathbf{H}^{H} \mathbf{H} \mathbf{S} \boldsymbol{\Phi}^{H} \mathbf{S} \right] \right) \\
        =& \dfrac{\alpha P}{P K + \sigma^2} N \operatorname{tr} \left( \mathbb{E}\left[\boldsymbol{\Phi}^{H} \mathbf{S} \boldsymbol{\Phi}^{H} \mathbf{S} \right] \right) \\
        =& 0 .
    \end{aligned}
\end{equation}
And the third term follows
\begin{equation}
    \begin{aligned}
        &\operatorname{tr} \left( \mathbb{E}\left[ \dfrac{\sqrt{\alpha P}}{P K + \sigma^2} \boldsymbol{\Phi}^{H} \boldsymbol{\Phi} \mathbf{Y}^{H} \mathbf{H} \mathbf{S} \right] \right) \\
        =& \dfrac{\sqrt{\alpha P}}{P K + \sigma^2} K \operatorname{tr} \left( \mathbb{E}\left[ \mathbf{Y}^{H} \mathbf{H} \mathbf{S} \right] \right) \\
        =& \dfrac{\sqrt{\alpha P}}{P K + \sigma^2} K \operatorname{tr} \left( \mathbb{E}\left[ \sqrt{(1-\alpha)P} \mathbf{S}^{H} \mathbf{H}^{H} \mathbf{H} \mathbf{S} \right] \right) \\
        =& \dfrac{\sqrt{\alpha (1-\alpha)} P}{P K + \sigma^2} K^2 L N .
    \end{aligned}
\end{equation}

\bibliographystyle{IEEEtran}
\bibliography{ref}

\begin{thebibliography}{10}
\providecommand{\url}[1]{#1}
\csname url@samestyle\endcsname
\providecommand{\newblock}{\relax}
\providecommand{\bibinfo}[2]{#2}
\providecommand{\BIBentrySTDinterwordspacing}{\spaceskip=0pt\relax}
\providecommand{\BIBentryALTinterwordstretchfactor}{4}
\providecommand{\BIBentryALTinterwordspacing}{\spaceskip=\fontdimen2\font plus
\BIBentryALTinterwordstretchfactor\fontdimen3\font minus \fontdimen4\font\relax}
\providecommand{\BIBforeignlanguage}[2]{{%
\expandafter\ifx\csname l@#1\endcsname\relax
\typeout{** WARNING: IEEEtran.bst: No hyphenation pattern has been}%
\typeout{** loaded for the language `#1'. Using the pattern for}%
\typeout{** the default language instead.}%
\else
\language=\csname l@#1\endcsname
\fi
#2}}
\providecommand{\BIBdecl}{\relax}
\BIBdecl

\bibitem{10492466}
T.~T.~T. Le \emph{et~al.}, ``A survey on random access protocols in direct-access leo satellite-based {IoT} communication,'' \emph{IEEE Commun. Surv. Tutorials}, vol.~27, no.~1, pp. 426--462, 2025.

\bibitem{10379539}
Z.~Wang \emph{et~al.}, ``A tutorial on extremely large-scale {MIMO} for {6G}: Fundamentals, signal processing, and applications,'' \emph{IEEE Commun. Surv. Tutorials}, vol.~26, no.~3, pp. 1560--1605, 2024.

\bibitem{svantesson_capacity_2005}
T.~Svantesson and B.~Rao, ``Capacity of spatio-temporally structured {MIMO} channels with estimation errors,'' in \emph{Proceedings of the IEEE International Conference on Acoustics, Speech, and Signal Processing (ICASSP 2005)}, Philadelphia, Pennsylvania, USA, 2005, pp. 401--404.

\bibitem{hassibi_how_2003}
B.~Hassibi and B.~Hochwald, ``How much training is needed in multiple-antenna wireless links?'' \emph{IEEE Trans. Inform. Theory}, vol.~49, no.~4, pp. 951--963, 2003.

\bibitem{coldrey_training-based_2007}
M.~Coldrey and P.~Bohlin, ``Training-based {MIMO} systems—part {I}: Performance comparison,'' \emph{IEEE Trans. Signal Process.}, vol.~55, no.~11, 2007.

\bibitem{ngo_energy_2013}
H.~Q. Ngo and T.~L. Marzetta, ``Energy and spectral efficiency of very large multiuser {MIMO} systems,'' \emph{IEEE Trans. Commun.}, vol.~61, no.~4, 2013.

\bibitem{liu_massive_2018}
L.~Liu and W.~Yu, ``Massive connectivity with massive mimo—part {I}: Device activity detection and channel estimation,'' \emph{IEEE Trans. Signal Process.}, vol.~66, no.~11, pp. 2933--2946, 2018.

\bibitem{yuan_fundamental_2018}
X.~Yuan, C.~Fan, and Y.~J. Zhang, ``Fundamental limits of training-based uplink multiuser {MIMO} systems,'' \emph{IEEE Trans. Wireless Commun.}, vol.~17, no.~11, pp. 7544--7558, 2018.

\bibitem{zhang_superimposed_2016}
H.~Zhang, S.~Gao, D.~Li, H.~Chen, and L.~Yang, ``On superimposed pilot for channel estimation in multicell multiuser {MIMO} uplink: Large system analysis,'' \emph{IEEE Trans. Veh. Technol.}, vol.~65, no.~3, pp. 1492--1505, 2016.

\bibitem{verenzuela_spectral_2018}
D.~Verenzuela and L.~Sanguinetti, ``Spectral and energy efficiency of superimposed pilots in uplink massive {MIMO},'' \emph{IEEE Trans. Wireless Commun.}, vol.~17, no.~11, 2018.

\bibitem{haritha_superimposed_2024}
H.~Haritha, D.~N. Amudala, R.~Budhiraja, and A.~K. Chaturvedi, ``Superimposed versus regular pilots for hardware impaired rician-faded cell-free massive {MIMO} systems,'' \emph{IEEE Trans. Commun.}, vol.~72, no.~11, pp. 6688--6706, 2024.

\bibitem{zhou_optimized_2024}
X.~Zhou, Y.~Zhu, W.~Xia, J.~Zhang, and K.-K. Wong, ``Optimized payload length and power allocation for generalized superimposed pilot in {URLLC} transmissions,'' \emph{IEEE Trans. Commun.}, vol.~72, no.~10, pp. 6073--6086, 2024.

\bibitem{duran_superimposed_2024}
M.~Duran, F.~Riera-Palou, G.~Femenias, H.~Q. Ngo, and M.~F.-G. García, ``Superimposed training in cell-free massive {MIMO}: Is it really worth it?'' in \emph{2024 IEEE 99th Vehicular Technology Conference (VTC2024-Spring)}, Singapore, Singapore, 2024, pp. 1--7.

\bibitem{zhou_power_2022}
X.~Zhou, W.~Xia, Q.~Zhang, J.~Zhang, and H.~Zhu, ``Power allocation of superimposed pilots for {URLLC} with short-packet transmission in {IIoT},'' \emph{IEEE Wireless Commun. Lett.}, vol.~11, no.~11, pp. 2365--2369, 2022.

\bibitem{4784839}
P.~Xia, S.-K. Yong, H.~Niu, J.~Oh, and C.~Ngo, ``Dft structured codebook design with finite alphabet for high speed wireless communication,'' in \emph{2009 6th IEEE Consumer Communications and Networking Conference}, 2009, pp. 1--5.

\end{thebibliography}

\end{document}